\documentclass{aa} 
\pdfoutput=1
\usepackage{graphicx}
\usepackage{txfonts}
\usepackage{natbib}
\begin{document}

   \title{WASP-94 A and B planets: hot-Jupiter cousins in a twin-star system\thanks{The radial-velocity and photometric data used for this work are only
   available in electronic form at the CDS via anonymous ftp to cdsarc.u-strasbg.fr (130.79.128.5) or via http://cdsweb.u-strasbg.fr/cgibin/qcat?J/A+A/}}
   
   \titlerunning{WASP-94 A \& B planets: Hot-Jupiter cousins}

   \author{M. Neveu-VanMalle\inst{1,2}
          \and
          D. Queloz\inst{2,1}
          \and
          D. R. Anderson\inst{3}
          \and
          C. Charbonnel\inst{1}
          \and
          A. Collier Cameron\inst{4}
          \and
          L. Delrez\inst{5}
          \and
          M. Gillon\inst{5}
          \and
          C. Hellier\inst{3}
          \and
          E. Jehin\inst{5}
          \and
          M. Lendl\inst{1,5}
          \and
          P. F. L. Maxted\inst{3}
          \and
          F. Pepe\inst{1}
          \and
          D. Pollacco\inst{6}
          \and
          D. S\'egransan\inst{1}
          \and
          B. Smalley\inst{3}
          \and
          A. M. S. Smith\inst{3,7}
          \and
          J. Southworth\inst{3}
          \and
          A. H. M. J. Triaud\inst{8,1}
          \and
          S. Udry\inst{1}
          \and
          R. G. West\inst{6}
          }

   \institute{Observatoire Astronomique de l'Universit\'e de Gen\`eve,
   Chemin des Maillettes 51, 1290 Sauverny, Switzerland
          \and
             Cavendish Laboratory, J J Thomson Avenue, Cambridge, CB3 0HE, UK
         \and
            Astrophysics Group, Keele University, Staffordshire, ST5 5BG, UK
         \and
            SUPA, School of Physics and Astronomy, University of St. Andrews, North Haugh,  Fife, KY16 9SS, UK
         \and
            Institut d'Astrophysique et de G\'eophysique, Universit\'e de
            Li\`ege, All\'ee du 6 Ao\^ut, 17, Bat. B5C, Li\`ege 1, Belgium
         \and
            Department of Physics, University of Warwick, Gibbet Hill Road, Coventry CV4 7AL, UK
         \and
            N.~Copernicus Astronomical Centre, Polish Academy of Sciences, Bartycka 18, 00-716 Warsaw, Poland
         \and
             Kavli Institute for Astrophysics \& Space Research,
             Massachusetts Institute of Technology, Cambridge, MA 02139, USA
                 }

   \date{Received 4 August, 2014; accepted 18 September, 2014}

% \abstract{}{}{}{}{}
% 5 {} token are mandatory

 \abstract
   {We report the discovery of two hot-Jupiter planets, each
orbiting one of the stars of a wide binary system.
   \object{WASP-94A} (\object{2MASS 20550794--3408079}) is an F8 type star hosting a transiting planet with a radius of 1.72
   $\pm$ 0.06 R$_{\rm Jup}$, a mass of 0.452 $\pm$ 0.034 M$_{\rm Jup}$ , and an orbital period of 3.95 days. The Rossiter--McLaughlin effect is clearly
   detected, and the measured projected spin--orbit angle indicates that the planet occupies a retrograde orbit.
   \object{WASP-94B} (\object{2MASS 20550915--3408078}) is an F9 stellar companion at an angular separation of 15\arcsec
   (projected separation 2700\,au),
   hosting a gas giant with a minimum mass of 0.618 $\pm$ 0.028 M$_{\rm Jup}$ with a period of 2.008 days, detected by
   Doppler measurements. 
   The orbital planes of the two planets are inclined relative to each other, indicating that at least one of them is inclined relative to the plane of the stellar binary.
   These hot Jupiters in a binary system bring new insights into
   the formation of close-in giant planets and the role of stellar multiplicity.}


   \maketitle
%
%________________________________________________________________

\section{Introduction}

More than 200 hot Jupiters \citep{Mayor:1995fk} have been discovered by exoplanet transit surveys. These short-period planetary systems are the easiest to detect because their small orbits improve their chances of transiting to typically 10\%. The large radii of hot Jupiters produce deep transits (of about 1\% depth) that are easily detectable from the ground. Several surveys have found these planets: TrES \citep{Alonso:2004qy}, XO \citep{McCullough:2005lr}, WASP \citep{Pollacco:2006qy}, HATNet \citep{Bakos:2007fj}, KELT \citep{Pepper:2007fk}, and QES \citep{Alsubai:2013uq}. Although
many hot Jupiters  are known, they rarely occur around solar-type stars with rates from 0.3 to 1.5\% \citep{Wright:2012lr} depending on the metallicity ([Fe/H]) of the stellar sample studied.

Only three stellar binaries are known to host pairs of circum-primary planetary systems. \object{HD20782}/\object{HD20781} is a wide stellar binary in which HD20782 hosts a Jupiter-mass planet \citep{Jones:2006lr} and HD20781 hosts two Neptune-mass planets \citep[announced by][]{Mayor:2011fk}. \object{XO-2} is a metal-rich wide stellar binary in which \object{XO-2N} hosts a transiting hot Jupiter \citep{Burke:2007uq}. \object{XO-2S} was found to have two planets, one of Jupiter, one of Saturn mass \citep{Desidera:2014lr}. \object{Kepler-132} is a stellar binary hosting three super-Earths \citep{Lissauer:2014lr}. Its angular separation of 0\farcs9 is too small to identify which star the planets are transiting. Dynamical considerations demonstrate that the two planets with the shortest periods cannot be orbiting the same star. We report here the discovery of a twin binary system in which each star hosts a hot Jupiter.

The WASP-94 unique system is of particular interest for understanding the formation of hot Jupiters. The conditions triggering the migration inside the disc \citep{Lin:1996lr, Papaloizou:2006lr} are not yet fully understood. Dynamical interactions with an additional companion \citep{Fabrycky:2007fk} are thought to play an important role, especially in the case of misaligned planets \cite[e.g.][]{Brown:2012}.

We here describe the planet detection processes, derive the system parameters, and discuss how this system fits into formation theories with respect to stellar multiplicity.

%__________________________________________________________________

\section{Discovery of WASP-94A b}

The WASP-94 system was identified as a transiting planet candidate by the WASP-South \citep{Hellier:2011kx} telescopes located in South Africa. A total of 18 200 photometric measurements were obtained by one of the eight WASP cameras between May 2006 and June 2012. WASP-94 was rapidly identified as a visual binary with an angular separation of 15\arcsec. Both stars fell in the photometric aperture (48\arcsec), making it impossible to distinguish which star was variable from WASP photometry alone.

\begin{figure}
   \centering
    \resizebox{\hsize}{!}{\includegraphics{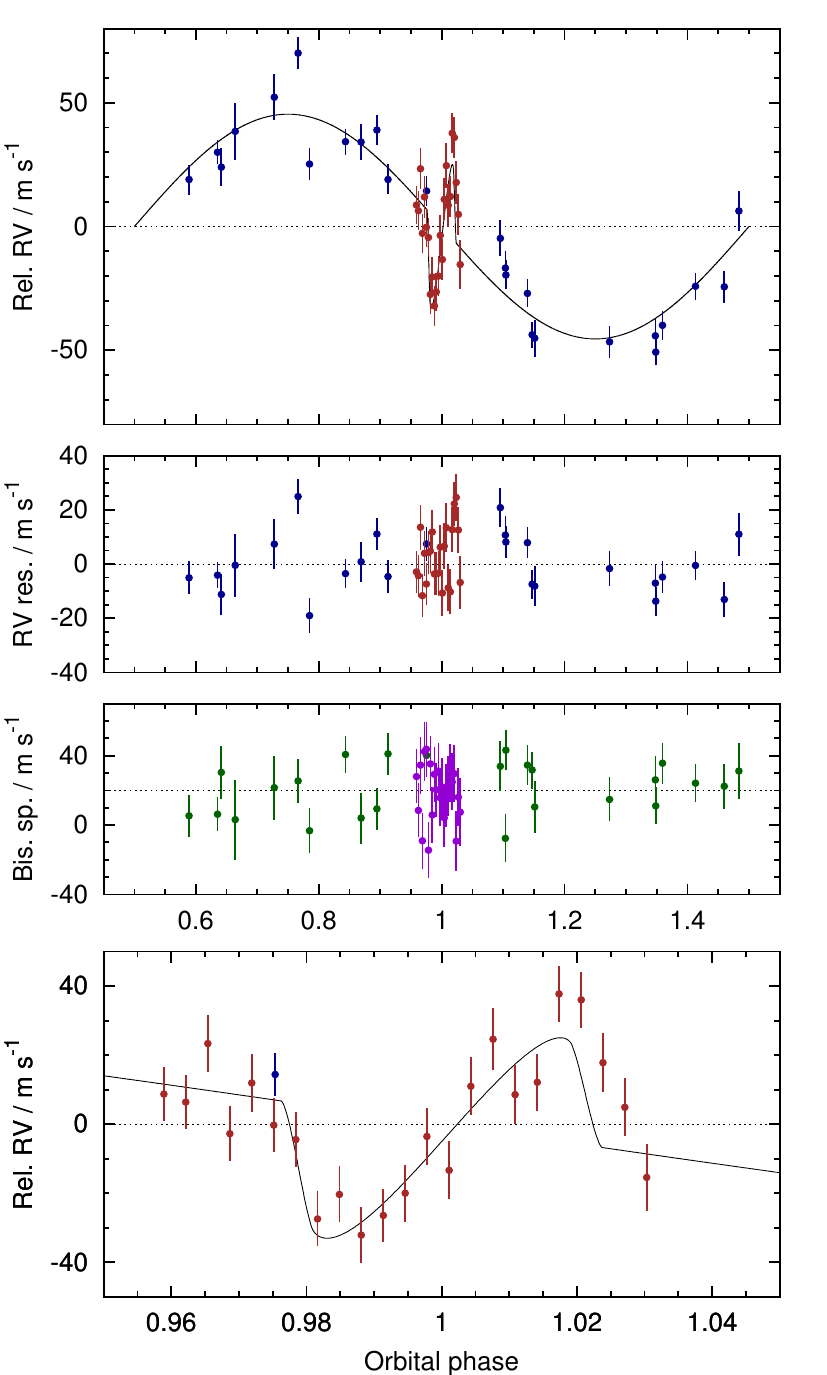}}
      \caption{\object{WASP-94Ab} spectroscopic data and fitted MCMC model. Top: the CORALIE radial velocities phased at the transit period over-plotted with the best-fit solution. Second panel: the residuals
      of the fit. Third panel: the bisector spans. Bottom: zoom on the CORALIE radial velocities taken during the transit over-plotted with the best solution for the Rossiter-McLaughlin effect.}
         \label{w94a_rv}
   \end{figure}

Observations of the brighter (western) component of the binary started in July 2012 with the spectrograph CORALIE \citep{Baranne:1996lr,Queloz:2000fk}. The radial velocities were computed by applying the weighted cross-correlation technique \citep{Pepe:2002qy} using a G2 mask. The transit-search algorithms of \citet{Collier-Cameron:2006qy} applied to the WASP photometry reveal a periodicity of 3.95\,d. A similar period was observed in the radial velocity data. No variability of the cross-correlation function (CCF) bisector spans was detected (see Fig.\,\ref{w94a_rv}). This confirms the planetary nature of the transiting candidate on the bright component \citep{Queloz:2001uq}. The unambiguous detection of the Rossiter-McLaughlin effect further confirms the planetary origin of the transit event. A total of 48 radial velocity (RV) measurements were collected during three seasons between 2012 Jul 16 and 2014 Jul 19. Half of them were taken during the transit on 2014 Jul 19 to measure the Rossiter-McLaughlin effect.

\begin{figure}
   \centering
   \resizebox{\hsize}{!}{\includegraphics{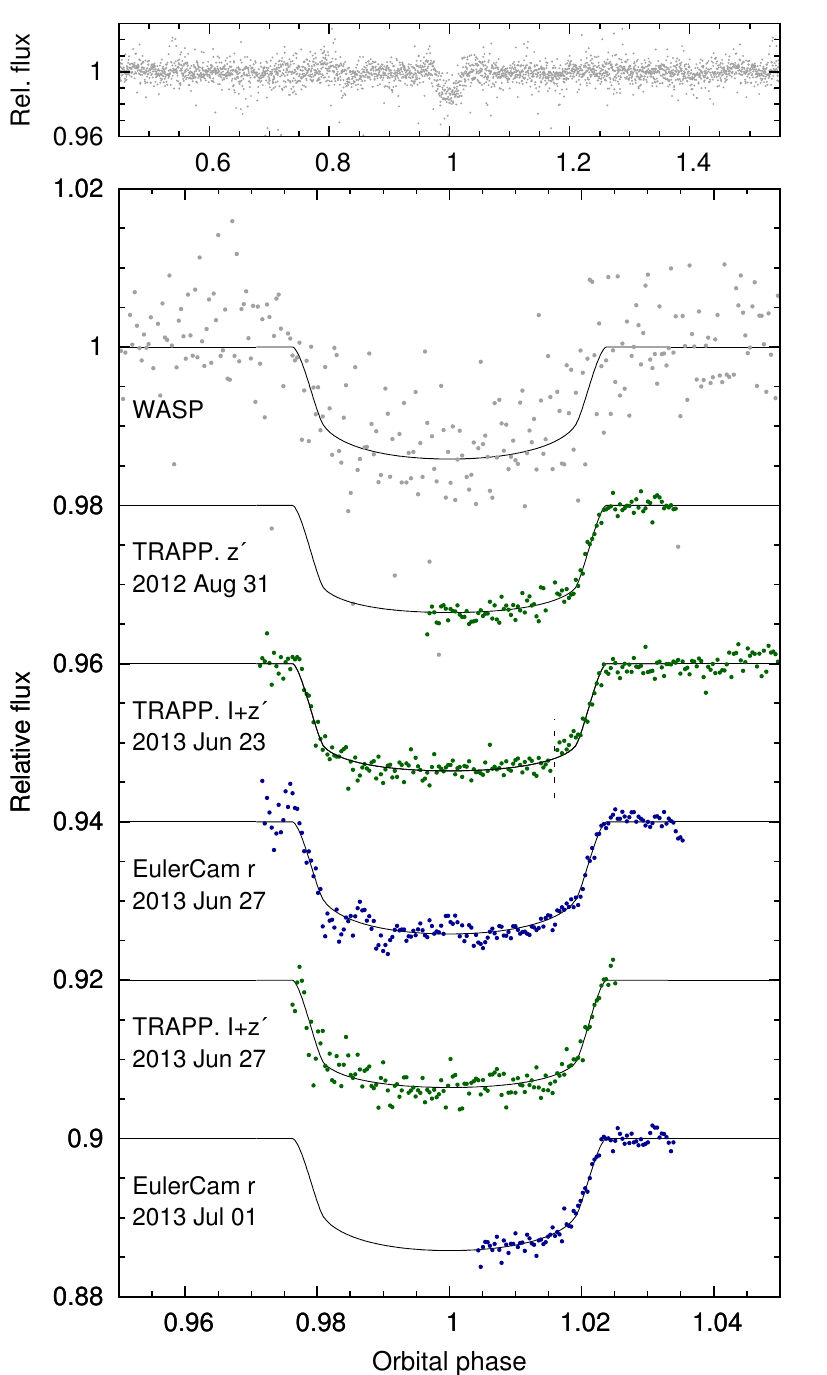}}
      \caption{WASP-94Ab photometric data and fitted MCMC model.
               Top: the phase-folded WASP data.
               Bottom: the binned WASP data and the follow-up high-precision transit light-curves.}
         \label{w94a_phot}
   \end{figure}

Following the confirmation of a planetary companion to WASP-94 with CORALIE, we used the TRAPPIST telescope \citep{Jehin:2011fj} to observe an egress in the $z$ band on 2012 Aug 31, which unambiguously confirmed that the transit occurrs on the brighter western component. Four high-precision transit light-curves were obtained in 2013. Two were observed with TRAPPIST in the $I + z$ filter on Jun 23 and 27. Simultaneously with TRAPPIST, EulerCam \citep{Lendl:2012lr} observed the full transit on Jun 27 in the Gunn $r'$ filter. A partial transit was observed by EulerCam on Jul 1, in the same filter. The first half of the transit on Jun 27 was affected by clouds (see Fig.\,\ref{w94a_phot}). The mean full-width at half-maxima (FWHM) of the point spread functions (PSF) of the observations are 3\farcs0, 3\farcs6 and 2\farcs7 for TRAPPIST and 3\farcs4 and 3\farcs3 for EulerCam. The separation of the two stars in this visual binary is 15\farcs0, therefore there is negligible contamination of the transit signal by the companion to WASP-94A.

We performed a spectral analysis on a high S/N spectrum obtained by co-adding all the individual CORALIE spectra. The methods used are described in \citet{Doyle:2013lr}. The spectral type was determined from $T_{\rm eff}$ using the stellar spectral classification of \citet{Gray:2008qy}. The stellar mass and radius were derived from the $ T_{\mathrm eff}$, $\log g,$ and {[Fe/H]} using the calibration reported
by \citet{Torres:2010fk}. A search for rotational modulation was performed as described by \citet{Maxted:2011lr} on the WASP photometry including the light from both stars, and led to no significant detection. The results are summarised in Table\,\ref{StPar}.

\begin{table}
   \centering
\caption{Stellar parameters for WASP-94A and B, including the results of the spectroscopic analysis.}
\small
\begin{tabular}{l  c c}
\hline\hline
Parameter & WASP-94A & WASP-94B\\
\hline
RA (J2000) & 20$^{\rm h}$55$^{\rm m}$07.94$^{\rm s}$ & 20$^{\rm h}$55$^{\rm m}$09.16$^{\rm s}$\\
Dec (J2000) & --34\degr08\arcmin07.9\arcsec & --34\degr08\arcmin07.9\arcsec\\
$V_{\rm mag}$ & 10.1 & 10.5\\
2MASS & 20550794--3408079 & 20550915--3408078\\
Rot.\ modulation & \multicolumn{2}{c}{$<$\,1 mmag (95\%)}\\
\hline
Spectral type & F8 & F9\\
$T_\mathrm{eff}$ (K)      & 6170 $\pm$ 80   & 6040 $\pm$ 90\\
$\log g$      & 4.27 $\pm$ 0.07 & 4.26 $\pm$ 0.06\\
$v\,\sin i_\star$ (km\,s$^{-1}$)     & 4.2 $\pm$ 0.5 & $<$1.5\\
{[Fe/H]}   &   +0.26 $\pm$ 0.15 & +0.23 $\pm$ 0.14\\
log A(Li)  &   2.10 $\pm$ 0.07 & $<$1.20\\
$M_\star$ (M$_{\rm \sun}$) & 1.29 $\pm$ 0.10 &  1.24 $\pm$ 0.09\\
$R_\star$ (R$_{\rm \sun}$) & 1.36 $\pm$ 0.13 &  1.35 $\pm$ 0.12\\
\hline
\end{tabular}
\label{StPar}
\end{table}

We performed a simultaneous Markov chain Monte Carlo (MCMC) least-squares fit to the combined radial velocities from CORALIE and the transit light-curves from WASP, TRAPPIST, and EulerCam. Details of the method were presented in \citet{Collier-Cameron:2007fk} and \citet{Anderson:2014lr}. The WASP data were scaled using the magnitudes found in the UCAC4 catalogue \citep{Zacharias:2013gf} to account for the dilution. We assumed a circular orbit because no significant eccentricity can be detected in the data ($e<\,0.13$ at $3\sigma$). We applied a prior on $v \sin i_{\star}$ corresponding to the projected equatorial rotation velocity of WASP-94A measured from stellar line broadening. The fitted value is consistent with the prior. To mitigate the effects of stellar activity on our results, the Rossiter-McLaughlin sequence was fitted as an independent dataset. The orbit is certainly misaligned and likely is retrograde ($\lambda=151^\circ\pm20^\circ$). Figure\,\ref{w94a_phot} shows the plots of the best solution for photometric data, Fig.\,\ref{w94a_rv} the same for spectroscopic data. The parameters are summarised in Table\,\ref{PlParA}.

\begin{table}
\centering
\caption{Parameters for WASP-94A b from the MCMC analysis.}
\small
\begin{tabular}{lclc}
\hline\hline
Parameter & Value & Parameter & Value\\
\hline
$P$ (d) & 3.9501907$^{(+44)}_{(-30)}$ & $M_\star$ (M$_{\rm \sun}$) & 1.45$^{(+9)}_{(-9)}$\\
$T_{\rm c}$ (HJD) & 2456416.40138$^{(+26)}_{(-26)}$ &  $R_\star$ (R$_{\rm \sun}$) & 1.62$^{(+5)}_{(-4)}$\\
$T_{\rm 14}$ (d) & 0.1870$^{(+9)}_{(-7)}$ & $\log g_\star$ (cgs) & 4.181$^{(+13)}_{(-17)}$ \\
$T_{\rm 12}=T_{\rm 34}$ (d) & 0.0189$^{(+9)}_{(-5)}$ & $\rho_\star$ ($\rho_{\rm \sun}$) &0.344$^{(+11)}_{(-20)}$\\
$\Delta F=R_{\rm P}^{2}$/R$_{*}^{2}$ & 0.01197$^{(+17)}_{(-15)}$ & $T_{\rm eff}$ (K) & 6153$^{(+75)}_{(-76)}$\\
$b$ & 0.17$^{(+9)}_{(-8)}$ & $M_{\rm P}$ (M$_{\rm Jup}$) & 0.452$^{(+35)}_{(-32)}$\\
$i$ (\degr) & 88.7$^{(+7)}_{(-7)}$ & $R_{\rm P}$ (R$_{\rm Jup}$) & 1.72$^{(+6)}_{(-5)}$\\
$K_{\rm 1}$ (km s$^{-1}$) & 0.0454$^{(+28)}_{(-27)}$ & $\log g_{\rm P}$ (cgs) & 2.542$^{(+30)}_{(-33)}$\\
$\gamma$ (km s$^{-1}$) & --8.82313$^{(+48)}_{(-46)}$ & $\rho_{\rm P}$ ($\rho_{\rm J}$) & 0.089$^{(+8)}_{(-8)}$\\
$e$ & $<$\,0.13 at 3$\sigma$ &  $T_{\rm P, A=0}$ (K) & 1604$^{(+25)}_{(-22)}$\\
$a$ (AU)  & 0.055$^{(+1)}_{(-1)}$ & $\lambda$ (\degr) & 151$^{(+16)}_{(-23)}$ \\[0.5mm] \hline
\multicolumn{4}{l}{Errors are 1$\sigma$; Limb-darkening coefficients were:}\\
\multicolumn{4}{l}{(Euler $r$) $a1$ =    0.630, $a2$ = --0.094, $a3$ =  0.478,
$a4$ = --0.276}\\
\multicolumn{4}{l}{(Trap $I+z$) $a1$ =    0.723, $a2$ = --0.450, $a3$ =  0.685,
$a4$ = --0.333}\\ \hline
\end{tabular}
\label{PlParA}
\end{table}

%__________________________________________________________________

\section{A planet around WASP-94B}
\label{sec:w94b}

The fainter (eastern) component of the stellar system, WASP-94B, was initially ignored during follow-up observations. From a couple of serendipitous measurements on WASP-94B, we detected a radial velocity variation of 159\,m\,s$^{-1}$. That triggered intensive observations of the star with CORALIE to confirm its variability. Interestingly, the FWHM of the CCF of WASP-94B is smaller than WASP-94A, suggesting a weak rotational broadening of the spectral lines ($v\,\sin i_\star$). Twenty-one spectra were obtained between 2012 Aug 18 and 2014 May 5. A spectral analysis was performed on WASP-94B similarly to WASP-94A. The results are shown in Table \,\ref{StPar}.

\begin{figure}
    \centering
    \resizebox{\hsize}{!}{\includegraphics{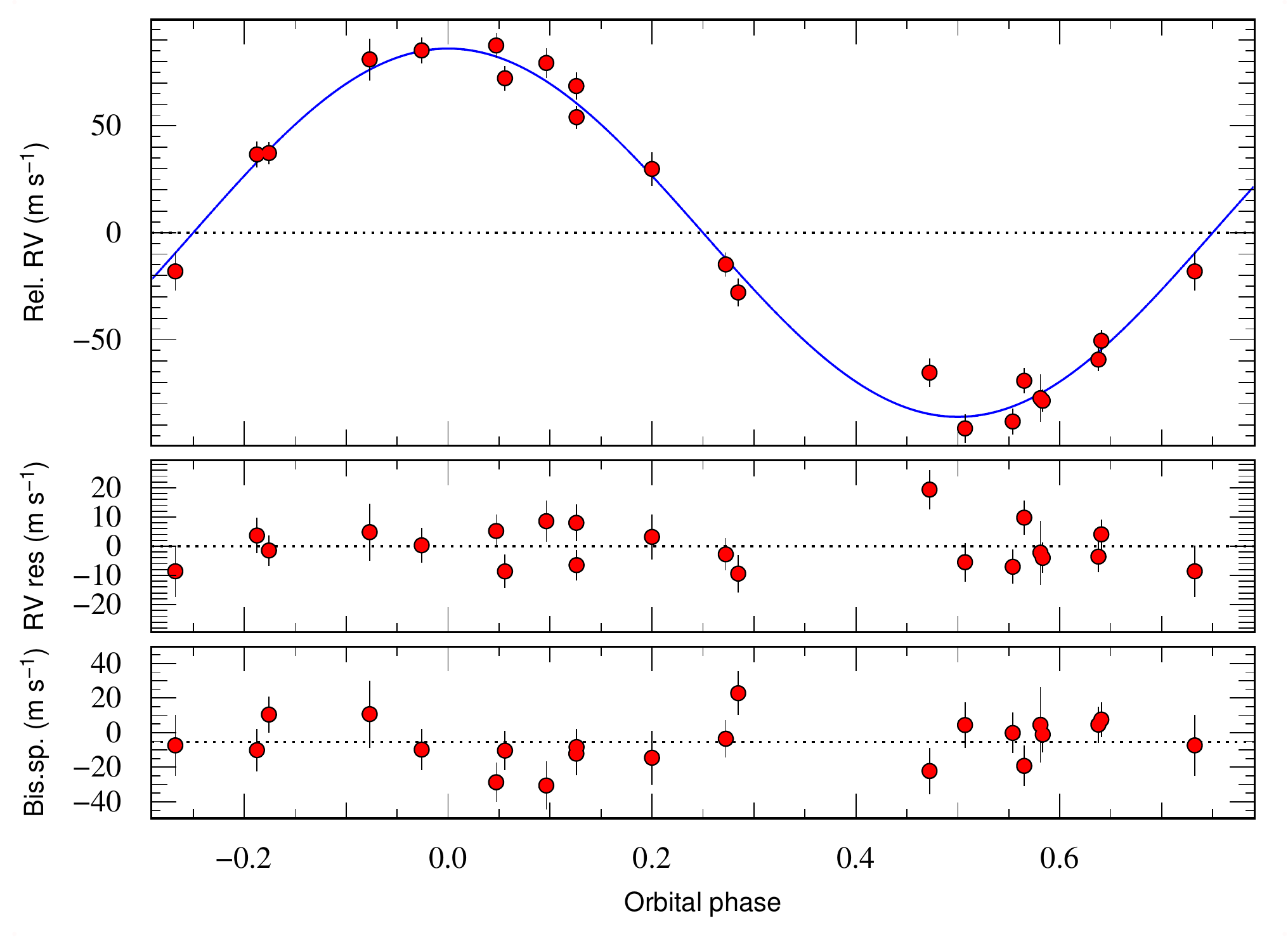}}
       \caption{WASP-94Bb radial velocities data. (Top) The phased radial velocities from CORALIE over-plotted with the best-fit solution.
       (Second panel) The residuals of the fit.
       (Bottom) The bisector spans.
               }
          \label{W94Bb_RV}
    \end{figure}

The FWHM and the bisector span of the CCF of the fainter component are scattered. Since there is no correlation with the radial velocities (see Fig.\,\ref{W94Bb_RV}), the possibility that stellar activity might have induced the signal is rejected \citep{Queloz:2001uq}.

We unsuccessfully searched for a transit of \object{WASP-94Bb}. First we used the WASP light-curve that includes the light of both stars. We phase-folded the WASP photometric data using the period and phase obtained with the radial velocities. No transit signal was detected. Later, the TRAPPIST telescope observed WASP-94B on 2014 May 10 and 14, excluding a transit down to 1.5 mmag depth. From this non-detection, we can estimate an upper limit for the inclination of the orbital plane of WASP-94Bb as i$\la$79\degr. It is interesting that the orbital planes of the two planets are inclined relative to each other, which indicates that at least one of them is inclined relative to the plane of the stellar binary.

\begin{figure}
    \centering
     \resizebox{\hsize}{!}{\includegraphics{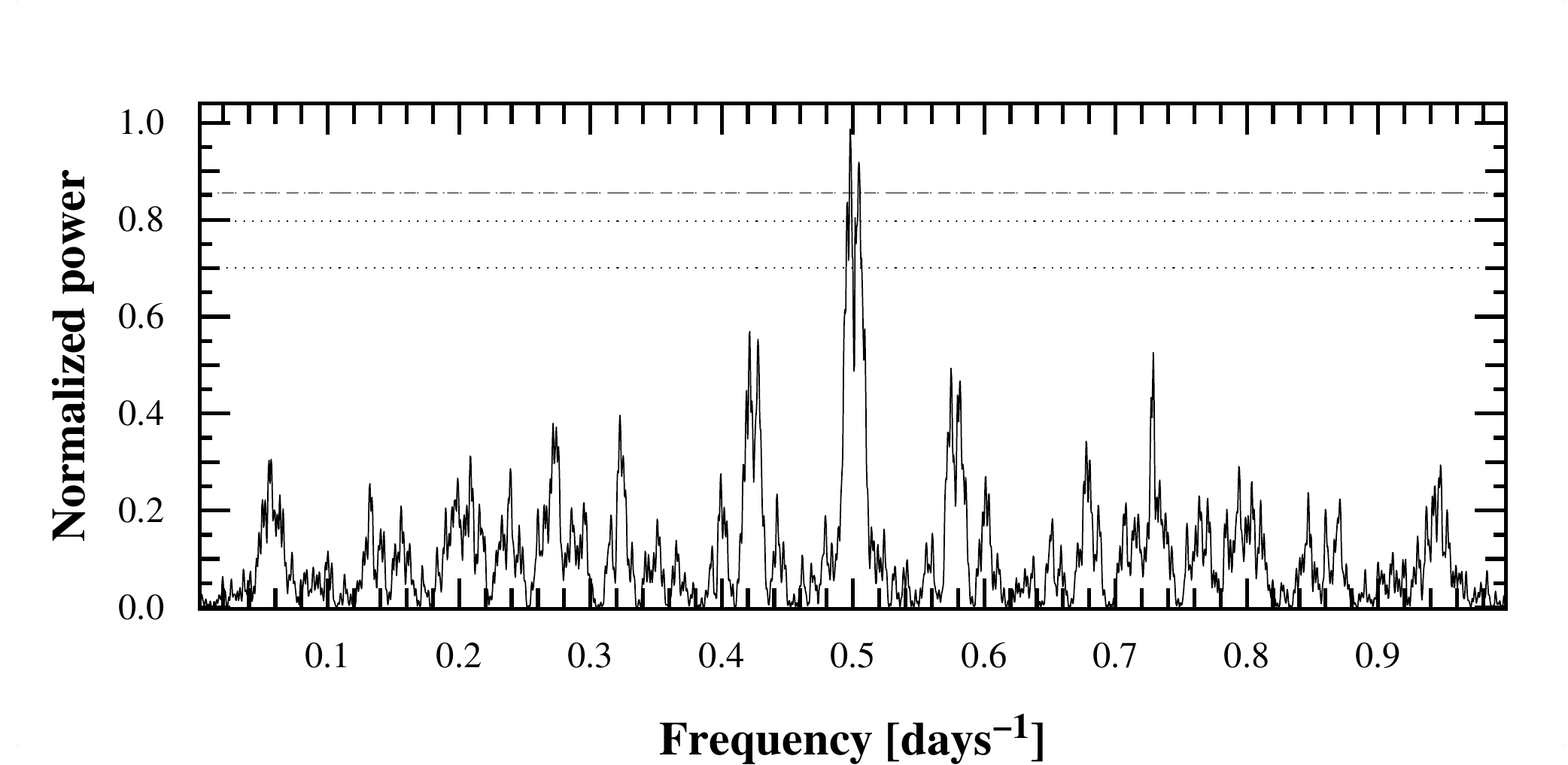}}
       \caption{WASP-94B periodogram. The dotted lines correspond to a false-alarm probability of 0.1\%, 1\% and 10\%.
               }
          \label{periofreq}
    \end{figure}

We only analysed the radial velocities because we did not detect any transit. The Lomb-Scargle periodogram of the velocity data plotted in frequency is displayed in Fig.\,\ref{periofreq}. There is a significant peak (false-alarm probability lower than 0.1\%) corresponding to a period of 2.008\,d. The fitted eccentricity is much lower than the error bar ($e=0.13\pm0.20$). Thus we assumed a circular orbit. The Keplerian solution is plotted in Fig.\,\ref{W94Bb_RV}. The system parameters are listed in Table\,\ref{PlParB}.

\begin{table}
\centering
\caption{Parameters for WASP-94B b from the single Keplerian orbital solution.}
\small
\begin{tabular}{lclc}
\hline\hline
Parameter & Value & Parameter & Value\\
\hline
$P$ (d) & 2.00839$^{(+24)}_{(-24)}$ & $M_{\rm P}$\,sin\,$i$ (M$_{\rm Jup}$)  & 0.618$^{(+28)}_{(-29)}$\\
$K_{\rm 1}$ (km s$^{-1}$) & 0.08648$^{(+265)}_{(-275)}$ & $a_1$\,sin\,$i$ (10$^{-3}$ AU)  & 0.01594$^{(+50)}_{(-49)}$\\
$e$ & 0 (adopted)  & $f_1(m)$ (10$^{-9}$ M$_{\rm \sun}$) & 0.134$^{(+13)}_{(-12)}$\\
$T_{\rm 0}$ (BJD) & 2456574.359$^{(+23)}_{(-22)}$ & $a$ (AU)  & 0.0335$^{(+6)}_{(-5)}$ \\
\hline
N$_{\rm mes}$ & 21 & $\sigma_{O-C}$ (m s$^{-1}$) & 7.16\\
$\Delta T$ (years) & 1.71\\
\hline
\end{tabular}
\label{PlParB}
\tablefoot{$T_{\rm 0}$ represents the time of largest amplitude of the radial velocities.}
\end{table}

%__________________________________________________________________

\section{ The stellar binary}
\label{sec:bin_sys}

The pair of stars is listed in the Washington Double Star Catalog \citep{Mason:2001fk} with the identifier 20551--3408. It was first observed in 1834 by Sir John F.\ W.\ Herschel \citep[`HJ 5234',][position angle $=$ 92\fdg5, separation $=$ 15\farcs0]{Herschel:1847uq}. The last published observation in 1999 \citep[][position angle $=$ 89\fdg6, separation $=$ 15\farcs06$\pm$0\farcs02]{Hartkopf:2013fj} reveals no significant change in position angle or separation. The proper motions provided in the UCAC4 catalogue are consistent within the error bars (see Table \ref{Comp}). Adding to that the very similar systemic velocity obtained with the radial velocity data, it is very likely that the visual binary is a bound system. The space motions (Table \ref{Comp}) are compatible with young disc stars \citep{Leggett:1992fk}.

The low rotation velocity and the lithium abundances of the two stars indicate that they have already undergone magnetic braking and lithium depletion along the main sequence. Large lithium variations are expected for F--G main sequence stars depending on their initial rotational velocity \citep[see][]{Charbonnel:2005lr, Melendez:2010fk}. Therefore we cannot use the lithium indicator to estimate an age difference, nor exclude that the two stars are coeval. The similar metallicity favours the idea of simultaneous formation. From evolutionary tracks we have estimated an age of $\sim$4\,Gy. Using the relation of \citet{Barnes:2007fk} together with a system age of 4 Gy, we found a rotation period of $\sim$19.0\,d for the brighter component (WASP-94A, F8), and $\sim$21.7\,d for the fainter (WASP-94B, F9). We compared these values with the measured rotation periods computed using the respective $v\,\sin i_\star$. We calculated that the brighter component (WASP-94A, P$_{\rm rot}=19.5$\,d) is perpendicular to the line of sight, while the fainter one (WASP-94B, P$_{\rm rot}>45.5$\,d) is inclined by $>$60 degrees. Note that gyrochronology is not best suited for stars of this age, especially because close-in giant planets were shown to have an influence on their host stars' angular momentum \citep{Lanza:2010qy, Poppenhaeger:2014uq}. These numbers therefore need to be used with caution. They are given as supporting information for dynamical considerations.

   \begin{table}
   \centering
      \caption[]{Comparison of WASP-94 A and B. Proper motions are taken from UCAC4, space motions are given with respect to the Sun, U pointing towards the Galactic anticentre.}
         \label{PropMot}
         \small
         \begin{tabular}{l c c}
            \hline\hline
            Star       &   WASP-94A & WASP-94B\\
            \hline
            $\mu_{RA}$ (mas/yr) & 24.0 $\pm$ 0.9 & 25.7 $\pm$ 0.8\\
            $\mu_{Dec}$ (mas/yr) & --41.7 $\pm$ 1.3 & --42.6 $\pm$ 1.3\\
            $\gamma$ (km/s) & --8.82313 $\pm$ 0.00005 & --8.7412 $\pm$ 0.002\\
            \hline
            $U$ (km/s) & 17.5  $\pm$ 1.3 & 18.3  $\pm$ 1.4 \\
            $V$ (km/s) & --35.4  $\pm$ 4.0 & --36.1  $\pm$ 4.0\\
            $W$ (km/s) & --14.3  $\pm$ 2.3 & --15.6  $\pm$ 2.4\\
            \hline
         \end{tabular}
    \label{Comp}
   \end{table}

From the absolute magnitudes we estimated a distance of 180 $\pm$ 20 pc. Gaussian fits to the PSF of the two stars on 300 EulerCam images give a projected separation of 15\farcs0297 $\pm$ 0\farcs0128 (14\farcs9877 $\pm$ 0\farcs0162 from TRAPPIST). The distance between the stars is estimated to be at least 2700\,au.

%__________________________________________________________________

\section{Discussion}

Spectroscopic and photometric surveys have similarly shown that hot Jupiters are rare objects \citep{Wright:2012lr}. Detecting two hot Jupiters in a system of two stars is therefore very unlikely. However, there is an important characteristic we must consider: the metallicity. Indeed, metal-rich stars are known to host more giant planets than metal-poor stars \citep{Gonzalez:1997lr, Santos:2004yq, Fischer:2005rt}. Considering this, the likelihood of finding a hot Jupiter around a metal-rich star is higher. No statistical study has focused on hot Jupiters around metal-rich stars. An estimate of this occurrence can be obtained using the results published in \citet{Fischer:2005rt}. Their sample contains 122 stars with a [Fe/H] between 0.25 and 0.5. We counted the number of hot Jupiters they detected around these stars. If hot Jupiters are defined as planets with a minimum mass higher than 0.3\,M$_{\rm Jup}$ and a period shorter than 10\,d, there are five such planets. This yields an occurrence rate of 4.10\%. If we relax the constraint of the lowest minimum mass to 0.2\,M$_{\rm Jup}$, we obtain two more planets and an occurrence rate of 5.74\%. We consider 5\% to be a good approximation.

If we assume that multiplicity and hot-Jupiter occurrence are independent, the chance of finding two hot Jupiters around two metal-rich stars ([Fe/H] $>$ 0.25) is $\sim$0.25\%. This value is an upper limit of the estimated occurrence rate for WASP-94A+B because the metallicities of these stars are [Fe/H] $\sim$ 0.25. With only one system, no statistical inferences can be made. Finding this system may have been entirely serendipitous. We examined the WASP sample for similar systems. WASP-77 \citep{Maxted:2013fj} and WASP-85 (Triaud et al.\ in prep) are close visual binaries in which one stellar component hosts a transiting hot Jupiter. The HARPS spectrograph at the ESO 3.6\,m telescope was used to identify the star in these binaries that hosts the transiting planet. Radial velocities were taken on both components. No variation suggesting the presence of a hot Jupiter around the companion stars was detected. WASP-64 \citep{Gillon:2013kx} is a wide binary containing a star hosting a transiting hot Jupiter. The radial velocity measurements obtained on the stellar companion reveal that it is a spectroscopic binary. Thus it is hard to conclusively demonstrate the presence or absence of a planetary companion. A few additional systems are currently being surveyed.

The formation of the planets in the WASP-94 system is unlikely to have been influenced by interactions between the discs because
of their large separation \citep{Duchene:2010ys}. Considering that the stars were formed simultaneously from the same cloud and are very similar, it is reasonable to assume that the discs had a very similar composition and/or properties. The discovery of a hot Jupiter around each star suggests that the same formation process took place and that similar favourable conditions boosted the migration of the planets.

On the other hand, secular interactions are thought to play an important role in the formation of hot Jupiters, even in the case of a distant stellar perturber \citep{Naoz:2012zr}. The secular timescales are compatible with the age of the system, under the condition that the planet's orbit reaches very high eccentricities. High mutual inclinations could lead to high eccentricities through the Kozai-Lidov mechanism \citep[e.g.][]{Petrovich:2014lr}. The results obtained in Sects. \ref{sec:w94b} and \ref{sec:bin_sys} are compatible with these requirements. In addition, the high inclination of the orbit of WASP-94Ab is compatible with a dynamical formation scenario. Note that misaligned systems are common around hot stars. With an effective temperature of 6170\,K, WASP-94A is on the threshold of patterns proposed by \citet{Schlaufman:2010fk} and \citet{Winn:2010lr}.

Even though at this stage nothing can be proven, there are recent dynamical theories relevant to this system. \citet{Moeckel:2012fr} described interactions in which a planet orbiting one component of a stellar binary can ``jump'' to the other star. If the two giant planets were formed around the same star, planet-planet scattering could have occurred. This would have pushed one of the planets near the host star and ejected the second one, which could then have been captured by the stellar companion. As we do not know the eccentricity of the stellar system, we can also consider the ``flipping mechanism'' described by \citet{Li:2014fk},
in which a coplanar system leads to very high eccentricities for the planet.

%__________________________________________________________________

\begin{acknowledgements}

M.\ Neveu-VanMalle thanks R.\ Mardling, N.\ Santos, and S.\ Hodgkin for fruitful discussions. WASP-South is hosted by the South African Astronomical Observatory and we are grateful for their ongoing support and assistance. Funding for WASP comes from consortium universities and from the UK's Science and Technology Facilities Council. TRAPPIST is funded by the Belgian Fund for Scientific Research (Fond National de la Recherche Scientifique, FNRS) under the grant FRFC 2.5.594.09.F, with the participation of the Swiss National Science Fundation (SNF). M.\ Gillon and E.\ Jehin are FNRS Research Associates. This work was supported by the European Research Council through the European Union's Seventh Framework Programme (FP7/2007-2013)/ERC grant agreement number 336480. L.\ Delrez acknowledges the support of the F.R.I.A.\ fund of the FNRS. A.\ H.\ M.\ J.\ Triaud received funding from the Swiss National Science Foundation under grant number P300P2-147773.

\end{acknowledgements}

%-------------------------------------------------------------------

\bibliographystyle{aa}
\bibliography{biblio}

\end{document}